# Statistical Modelling and Analysis of the Computer-Simulated Datasets

M. Harshvardhan
*Indian Institute of Management Indore, India*

Pritam Ranjan
*Indian Institute of Management Indore, India*

**ABSTRACT**

*Over the last two decades, the science has come a long way from relying on only physical experiments and observations to experimentation using computer simulators. This chapter focusses on the modelling and analysis of data arising from computer simulators. It turns out that traditional statistical metamodels are often not very useful for analyzing such datasets. For deterministic computer simulators, the realizations of Gaussian Process (GP) models are commonly used for fitting a surrogate statistical metamodel of the simulator output. The chapter starts with a quick review of the standard GP based statistical surrogate model. The chapter also emphasizes on the numerical instability due to near-singularity of the spatial correlation structure in the GP model fitting process. The authors also present a few generalizations of the GP model, reviews methods and algorithms specifically developed for analyzing big data obtained from computer model runs, and reviews the popular analysis goals of such computer experiments. A few real-life computer simulators are also briefly outlined here.*

## 1. INTRODUCTION

In early days, when the computers were not readily accessible to common people, statisticians and data analysts focussed on the development of innovative methodologies that were efficient for analyzing small datasets. Over the last two decades, we have come a long way from relying on only physical experiments and observations to experimentation using computer simulation models, commonly referred to as the computer simulators or computer models. These simulators are software implementation of the real-world processes, imitated based on the comprehensive understanding on the underlying phenomena. The applications range from simulating socioeconomic behaviour, impact due to a car crash, manufacturing a compound for drug discovery, climate and weather forecasting, population growth of certain pest species, cosmological phenomena like dark energy and universe expansion, emulation of tidal flow for harnessing renewable energy, the simulation of a nuclear reactions, and so on. Given the easier access to high performance computing power such as cloud computing and cluster grids, computer model data is now a reality in everyday life.

In this chapter, we focus on the modelling and analysis of data sets arising from such computer simulators. Similar to the physical experiments setup, the data obtained from the computer simulator runs have to be modelled and analysed for a deeper understanding of the underlying process. However, traditional statistical metamodels are often not very useful for analyzing such datasets. This is because, many a time, these computer models are deterministic in nature, that is, the repeated runs of such a computer simulator with a fixed input settings yield the same output / response. In other words, there is no replication error for the deterministic computer simulators. Recall that in the traditional statistical models, such as regression, the main driving force for model fitting and inference part of methodology is the distribution of replication errors.

For deterministic computer simulators, the realizations of Gaussian Process (GP) models, trained by the observed simulator data, are commonly used for fitting a surrogate statistical metamodel of the simulator output. This is particularly crucial if the simulator is expensive to run, which is the case for many complex real-life phenomena. The notion of GP models gained popularity in late 1990 and early 2000 (e.g., Santner et al. (2003); Rasmussen and Williams (2006); Fang et al. (2005)), though it was first proposed in the seminal paper of Sacks et al. (1989). Section 2 of the chapter presents a quick review of the standard GP based statistical surrogate model. We will also briefly discuss the implementation procedure using both the maximum likelihood method and the Bayesian approach.

Almost all published research articles and books focus on the new methodologies and algorithms that can be used for analyzing the computer simulator data, and not on the small nuances related to the actual implementation which is extremely useful from a practitioners' standpoint. This chapter emphasizes on such computational issues. In particular, Section 3 of the chapter discusses the numerical instability due to near-singularity or ill-conditioning of the spatial correlation structure which is the key building block behind the flexibility of the GP-based surrogate model. In practice, the majority of researchers simply use a numerical fix to overcome this issue, but this inadvertently compromises with other aspects of the model assumptions. We present an empirical study to compare different current practices to address this ill-conditioning problem. We also discuss the best coding practices in the implementation of such model fitting exercise, for instance, which of the matrix decomposition method, LU / QR / SVD / Cholesky, is recommended from an accuracy and time efficiency perspective.

Given the revolution in the computing power, it is now easy to collect and process data sets that are spatio-temporal and functional in nature. Dynamic computer models, i.e. the simulator which returns time-series response (see Zhang et al. (2018b)), is a current hot topic of research in applied statistics and computer experiments. Section 4 of the chapter reviews several generalizations of the GP model that accounts for multiple sources of uncertainty in the simulation model, non-stationarity of the underlying processes, and dynamic nature of such computer simulator outputs.

With the advent of inexpensive high performance computing facilities on cloud servers and different grids, a plethora of big data is now available in the public domain. The standard methodologies and algorithms are typically not very efficient in analyzing such datasets. Section 5 of the chapter reviews methods and algorithms specifically developed for analyzing big data obtained from computer model runs. Some of the approaches are methodolgoical in nature, and use sparse matrix computation and localized model approximation based ideas to efficiently build the statistical surrogate, whereas others emphasize on the clever use of parallelization on CPUs and graphical processing units (GPUs) for handling the big data.

Section 6 of the chapter reviews the popular analysis goals of such computer experiments. For instance, Jones et al. (1998) proposed an innovative merit based criterion called the expected improvement for the process optimization; Linkletter et al. (2006) developed a variable screening approach for the identification of important inputs to the computer simulator and subsequently ignoring the non-important ones; Vernon et al. (2010), Pratola et al. (2013) and Ranjan et al. (2016) discussed the calibration of computer simulators to ensure the generation of realistic outputs. Finally, Section 7 presents brief outlines of a few real-life computer models.

Over the past decade or so, a few open source software (mostly in R) have been published which are becoming increasingly popular among the researchers and practitioners, for instance, GPfit (MacDonald et al., 2015), mlegp (Dancik and Dorman, 2008), TGP (Gramacy, 2007), DiceKriging (Roustant et al., 2012), laGP (Gramacy et al., 2016) and DynamicGP (Zhang et al., 2018a). In this chapter, we use several test function based computer simulators and real-life applications to illustrate the concepts and methodologies via these packages. We also provide code snippets of R to help understand how to apply use them in your research endeavours.

## 2. GAUSSIAN PROCESS MODEL

A stochastic process is a collection of random variables indexed by time or space. A Gaussian process is commonly used in statistical modelling because of its nice distributional properties and closed form expressions of moments and other summary statistics. In notation, $\{z(x), x \in [0,1]^d\}$, in short, $z(x) \sim GP(0, \sigma_z^2 R)$ with $E(z(x)) = 0$, $Var(z(x)) = \sigma_z^2$, and $Cov(z(x_i), z(x_j)) =$ $\sigma_z^2 R(x_i, x_j)$ where $R$ is a positive definite correlation function. Then, any finite subset of variables $\{z(x_1), z(x_2), \ldots, z(x_n)\}$, for $n \geq 1$, jointly follows multivariate normal distribution.

In conventional regression models, we set $y_i = f(x_i, \beta) + \varepsilon_i$, where $\varepsilon_i$'s are i.i.d. $N(0, \sigma^2)$. Though the regression model can be very flexible if we choose the $f(x_i, \beta)$ carefully, this is not suitable for emulating the deterministic computer model outputs, as there is no replication error. In GP model (also sometimes referred to as the GP regression model), we aim to find a surrogate that is an interpolator of all the observed training data, that is, the fitted surface passes through all original $(x_i, y_i), i = 1, 2, \ldots, n$. In between the training points, the smoothness and curvature of the fitted surrogate is guided by the correlation structure $R(\cdot, \cdot)$. The GP model is formally presented in the next subsection.

## 2.1 Model Statement

Let the $i$-th $d$-dimensional input and 1-dimensional output of the computer simulator be denoted by $x_i = (x_{i1}, x_{i2}, \ldots, x_{id})$ and $y_i = y(x_i)$, respectively. Suppose the set of all $n$ training data are held together in the design $D = \{x_1, x_2, \ldots, x_n\}$ and the output vector $Y = (y_1, y_2, \ldots, y_n)'$. Then, the GP model is written as

$$y_i = \mu + z(x_i), \quad i = 1, 2, \ldots, n, \tag{1}$$

where $\mu$ is the overall mean and $z(x) \sim GP(0, \sigma_z^2 R)$. Subsequently, $Y$ follows multivariate normal distribution with mean $1_n \mu$ and variance-covariance matrix $\Sigma = \sigma_z^2 R_n$, where $1_n$ is an $n \times 1$ vector of all 1's, and $R_n$ is an $n \times n$ correlation matrix with $(i, j)$-th element given by $R(x_i, x_j)$ (see Sacks et al. (1989); Santner et al. (2003) for more details).

The most crucial component of this GP model is the correlation structure, which dictates the 'smoothness' of the interpolator that passes through the observations. In a multidimensional scenario, it tells us how wobbly and differentiable the fitted surrogate is. By definition, any positive definite correlation structure would suffice, but the most popular choice is the Gaussian correlation. In Machine Learning and Geostatistics literature, Gaussian correlation is also referred to as the radial basis function. Gaussian correlation is a special case (with $p_k = 2$) of the power-exponential correlation given by

$$R(x_i, x_j) = \prod_{k=1}^{d} \exp\{-\theta_k |x_{ik} - x_{jk}|^{p_k}\}, \tag{2}$$

where $\theta_k$ and $p_k$ controls the wobbliness of the surrogate in the $k$-th coordinate.

The model described by (1) and (2) is typically fitted by either maximizing the likelihood or via Bayesian algorithms like Markov chain Monte Carlo (MCMC). As a result, the predicted response $\hat{y}(x_0)$ for an arbitrary input $x_0$ can be obtained as a conditional expectation from the following $(n + 1)$-dimensional multivariate normal distribution:

$$\begin{pmatrix} y(x_0) \\ Y \end{pmatrix} = N\left( \begin{pmatrix} \mu \\ \mu 1_n \end{pmatrix}, \begin{pmatrix} \sigma_z^2 & \sigma_z^2 r'(x_0) \\ \sigma_z^2 r(x_0) & \sigma_z^2 R_n \end{pmatrix} \right), \quad (3)$$

where $r(x_0) = [corr(x_1, x_0), \ldots, corr(x_n, x_0)]'$. The predicted response $\hat{y}(x_0)$, which is also the best linear unbiased predictor (BLUP), is the same as the conditional mean:

$$E(y(x_0)|Y) = \mu + r(x_0)' R_n^{-1}(Y - 1_n \mu), \quad (4)$$

and the associate prediction uncertainty estimate (denoted by $s^2(x_0)$) can be quantified by the conditional variance:

$$Var(y(x_0)|Y) = \sigma_z^2 (1 - r'(x_0) R_n^{-1} r(x_0)). \quad (5)$$

In practice, the parameters $\mu, \sigma$ and $\theta = (\theta_1, \ldots, \theta_d)$ are replaced by their estimates (maximum likelihood estimates or posterior means in MCMC) in (4) and (5).

## 2.2 Implementation Details

The key aspects of the implementation here is to efficiently maximize the likelihood and evaluate the predicted mean response and associated uncertainty measure. For this GP model, the likelihood is simply the joint probability density function of the multivariate normal distribution of $Y$, i.e.,

$$-2\log(L) \propto \log(|R_n|) + n\log(\sigma_z^2) + \frac{(Y - 1_n \mu)' R_n^{-1}(Y - 1_n \mu)}{\sigma_z^2}, \quad (6)$$

where $|R_n|$ is the determinant of the $n \times n$ correlation matrix $R_n$.

Minimizing $-2\log(L)$ gives closed form expressions for $\hat{\mu}$ and $\hat{\sigma}_z^2$ as

$$\hat{\mu} = (1_{n'} R_n^{-1} 1_n)^{-1} (1_{n'} R_n^{-1} Y), \quad (7)$$

and

$$\hat{\sigma}_z^2 = \frac{(Y - 1_n \hat{\mu})' R_n^{-1}(Y - 1_n \hat{\mu})}{n}, \quad (8)$$

where $R_n$ is a function of unknown $\theta = (\theta_1, \ldots, \theta_d)$. Finding good estimates of the $d$-dimensional correlation hyperparameter vector $\theta$ is not easy. It is common to use numerical optimization techniques like multi-start Gauss-Newton type methods or evolutionary algorithms like particle swarm method and genetic algorithms to find $\hat{\theta}$.

For the convenience of researchers and practitioners in this area, several R packages have been developed that provide easy implementation of fitting this GP model, for example, TGP (Gramacy, 2007), mlegp (Dancik and Dorman, 2008), DiceKriging (Roustant et al., 2012) and GPfit (MacDonald et al., 2015).

In this section, we briefly illustrate the usage of GPfit (MacDonald et al., 2015) for fitting a GP model to a simulated data set. Suppose the simulator output is generated by a one-dimensional test function $f(x) = \log(x + 0.1) + \sin(5\pi x)$, and $X = \{x_1, \ldots, x_{10}\}$ is a randomly generated training set as per the space-filling Latin hypercube design. Then the GP model can be fitted using the following code:

```
GPmodel = GPfit::GP_fit(X, Y, corr = list(type="exponential", power=2))
```

The GPfit object GPmodel contains the parameter estimates, which can be further passed on for generating the predictions along with uncertainty estimates at a test set. Figure 1 shows the fitted surrogate along with the true response.

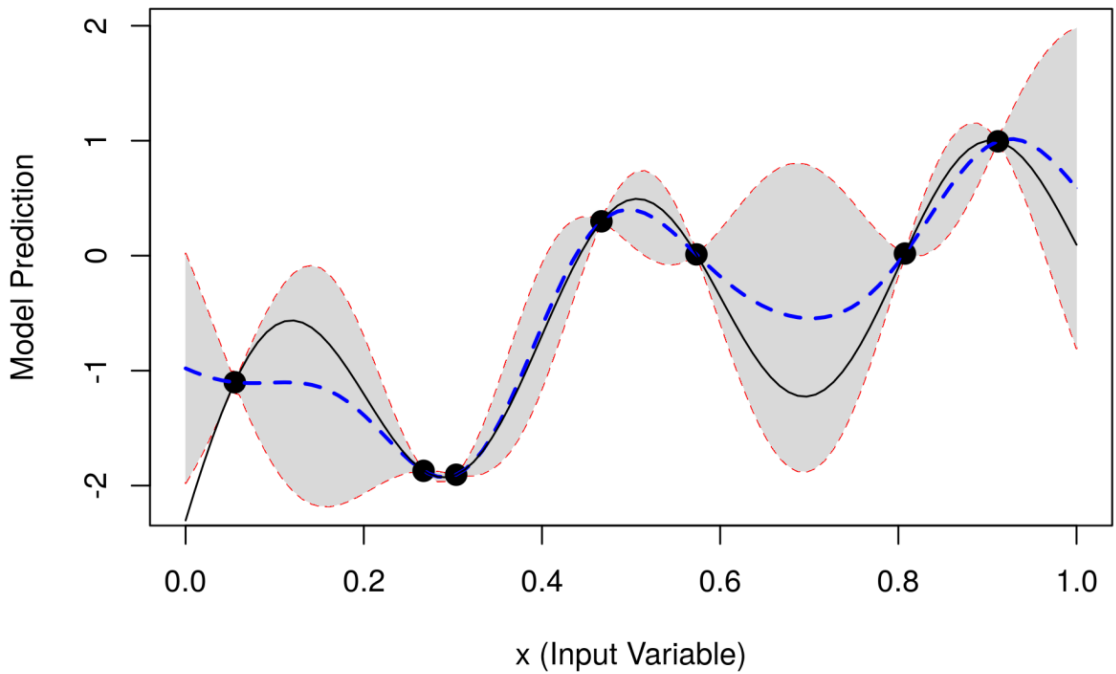

*Figure 1 The blue dashed curve is the mean prediction obtained using $GPfit$, the black solid curve is the true simulator response curve $f(x) = log(x + 0.1) + sin(5\pi x)$, the black solid dots are the training data points, and the shaded area represent the uncertainty quantification via $\hat{y}(x) \pm 2s(x)$.*

In GPfit package, the estimate of $\theta$ is obtained by minimizing the deviance $(-2\log(L_p)$, where $L_p$ is the profiled likelihood obtained after substituting $\hat{\mu}$ and $\hat{\sigma}_z^2$) using a multi-start gradient based search (L-BFGS-B) algorithm. As a side note, they use a slightly different parametrization, i.e., $\theta_k = 10^{\beta_k}$, and then find optimal $\beta = (\beta_1, \ldots, \beta_d)$ (see Section 3.4 for more discussion on reparametrization of $R(x_i, x_j)$). The starting points of L-BFGS-B are selected using the $k$-means clustering algorithm on a large space-filling design over the search space, after discarding $\beta$ vectors with high deviance. The control parameter is a vector of three tunable parameters used in the deviance optimization algorithm. The default values correspond to choosing $2d$ clusters based on $80d$ best points (smallest deviance) from a $200d$ - point random space-filling design in $\beta$-space. One can enhance the robustness of the optimal $\beta$ estimates by increasing the arguments of control in GP_fit, however, this is a computationally expensive, with $O(n^3)$ complexity, where $n$ is the size of the training data. Thus, one should balance between the computational cost and the robustness of likelihood optimization. For details see MacDonald et al. (2015).

## 3. COMPUTATIONAL ISSUES IN FITTING GP MODELS

Though fitting a GP model to the training data and prediction on a test set may seem like straightforward tasks, there are several issues like numerical instability, prediction accuracy, biases due to miss-specified model, and some concern due to the heavy computational cost, particularly when dealing with big data. In this section, we review a few such outstanding issues and popular approaches to address them.

### 3.1 Matrix Decomposition in Likelihood Evaluation

Different components of the GP model, including the likelihood (equivalently, the deviance expression), estimates of $\mu$ and $\sigma_z^2$, the predicted mean response and the uncertainty estimate (as shown in (4) - (8)), contain two computationally expensive terms, the determinant of $R_n$ and the inverse of $R_n$. For finding optimal $\theta$ (or equivalently, $\beta$, as in GPfit), these expressions have to be evaluated hundreds to thousands of times for different realizations of $\theta$. If the size of the training data, $n$, is small, numerous evaluations of $|R_n|$ and $R_n^{-1}$ by any method is not a concern from computational cost standpoint, however, for large $n$, fast evaluations of $|R_n|$ and $R_n^{-1}$ become crucial.

It is common to use matrix decomposition methods like LU, Cholesky, QR and SVD, for efficient computation of determinants and inverses of $R_n$, and terms like $R_n^{-1}w$, for some $n \times 1$ vector $w$. It turns out that these decomposition methods have different computational cost, and more importantly, exhibit different precision as well. In this section, we present a simulation study based comparison of these matrix decomposition methods for Gaussian correlation (2). The objective is to choose the right matrix decomposition method while implementing the GP model procedure.

The results are averaged over 1000 simulations. For each replication, we randomly generate $X = \{x_1, x_2, \ldots, x_n\}$ using a space-filling Latin hypercube design over $[0,1]^d$ and $\theta \in (0, \infty)^d$, and then evaluate $R_n$ as in (2). Subsequently, we perform the decomposition and then obtain the reconstituted matrix. For instance, for LU decomposition, we obtain the triangular matrices $L$ and $U$ via $\text{lu}(R_n)$, and then find $R_n^* = LU$. In theory, $R_n^* = R_n$, but in practice, they can be somewhat different. Both the empirical simulation study and the theoretical complexity measured in terms of big O, show that Cholesky and SVD have much greater accuracy and are computationally cheaper as compared to LU and QR.

Note that Cholesky decomposition method uses two sets of linear solves for computing $R_n^{-1}w$. That is, if $R_n = LL^T$, then $R_n^{-1}w = solve(L^T, solve(L, w))$. Whereas, the SVD method finds $R_n^{-1}$ by inverting the singular values as

$$R_n^{-1} = \sum_{i=1}^n u_i v_i^T / d_i,$$

where $R_n = UDV^T$ with $U = [u_1, \ldots, u_n], V = [v_1, \ldots, v_n]$ and $D = diag(d_1, \ldots, d_n)$. It turns out that for applications with large $n$ and small input dimension $d$, both of these matrix decomposition methods suffer from numerical instability due to ill-conditioning of $R_n$. In this chapter, we only focus on Cholesky and SVD as the other decomposition methods are less efficient and inaccurate.

## 3.2 Near-singularity of Correlation Matrix

Recall that an $n \times n$ matrix is said to be singular if at least one of its rows (or columns) is linearly dependent on the rest of rows (or columns), i.e., the matrix does not have full row (or column) rank, i.e., the determinant is zero. However in a near-singular matrix, the determinant is not exactly equal to zero but very small. One popular method of quantifying the near-singularity of $R_n$ is via its condition number defined by, $\kappa(R_n) = \| R_n^{-1} \| \cdot \| R_n \| = \lambda_n / \lambda_1$, where $\|\cdot\|$ is the $L_2$ norm of the matrix, and $\lambda_i$ is the $i$-th smallest eigen value of $R_n$. An $n \times n$ matrix $R_n$ is said to be near-singular (or ill-conditioned) if $\kappa(R_n)$ is large. For Gaussian correlation, the near-singularity occurs when $\sum_{k=1}^d \theta_k |x_{ik} - x_{jk}|^2 \approx 0$, which implies either the two data points $x_i$ and $x_j$ are close to each other and/or $\theta_k$'s are close to zero. This further implies that the condition number is directly proportional to the sample size $n$ and inversely proportional to $d$ and $\theta$. In other words, the larger the training data size, it is more likely to run into near-singularity, whereas if the input dimension and/or $\theta$ are large, it is less likely to run into near-singularity.

From an implementation standpoint, if the condition number is larger than say $1/\varepsilon_M$,

where $\varepsilon_M$ is the machine precision ($\varepsilon_M$=2.220446e-16 for our desktop computer), the determinant of $R_n$ would be too close to zero, and the linear solves using $R_n$ become too sensitive and unreliable, if at all obtainable. If $\kappa(R_n) > 1/\varepsilon_M$, Cholesky decomposition of $R_n$ would crash, rendering the method infeasible. However, one can use SVD approach and approximates $R_n^{-1}$ as

$$R_n^{-1} \approx \sum_{i=1}^{n} u_i v_i^T / d_i \cdot I(d_i > \eta),$$

where $\eta$ is a pre-specified threshold that determines a cut-off for not using very small singular values in approximating the inverse of $R_n$. For details, see Jones et al. (1998) and Booker et al. (1999). It turns out that the SVD based approach is very sensitive with respect to the choice of $\eta$. That is, a large value of $\eta$ would make the approximated $R_n^{-1}$ too far from the true (unobservable) $R_n^{-1}$, whereas a small value of $\eta$ would make approximated $R_n^{-1}$ unreliable due to the inclusion of very large $1/d_i$.

A popular technique to resolve this numerical issue is to use a "nugget" (or also referred to as a "jitter") term $\delta$ in the model by replacing $R_n^{-1}$ with $R_{n,\delta}^{-1}$, where $R_{n,\delta} = R_n + \delta I_n$ (Neal, 1997). This method works because $\kappa(R_{n,\delta}) = (\lambda_n + \delta)/(\lambda_1 + \delta)$ would be much smaller than $\kappa(R_n)$. Similar to the SVD based approximation, here also one needs to find $\delta$, but interestingly, this nugget based approach is less sensitive to the choice to $\delta$ as compared to selecting appropriate $\eta$ in the SVD method. Gramacy and Lee (2012) suggests estimating $\delta$ along with other model parameters in a Bayesian framework, however, the search space for $\delta$ has to be carefully chosen so that the lower limit is large enough to ensure well-conditioned $R_{n,\delta}$. To this effect, Ranjan et al. (2011) developed a lower-bound on $\delta$ which suffices well-behaved and accurate approximation of $R_n$.

If we choose $\eta, \delta > 0$, the resultant mean prediction function is not an interpolator. Thus, both the nugget and SVD based approaches lead to methodological consequences which may not be desirable for a deterministic simulator. Ranjan et al. (2011) proposed an iterative scheme that uses the lower bound of nugget to start with for well-behaved $R_{n,\delta}$ and then the iterative regularization makes the predictor converge to the interpolator. The following R code snippet illustrates the usage of GPfit package to specify the number of iteration (say, $M = 5$) in this iterative procedure:

```
GPprediction = GPfit::predict.GP(GPmodel, xnew, M=5)
```

Of course, one can argue on a philosophical ground that none of the realistic computer model is deterministic, and some sort of uncertainties and biases are always present. Thus, one must include a non-zero nugget term, and some amount of smoothing is a desirable feature for a predicted surrogate. Even in such a case, if the nugget parameter is estimated using the maximum likelihood method or a Bayesian approach (via MCMC), the lower limit of the search space for $\delta$ must be large enough to ensure well-conditioned $R_{n,\delta}$, for which the lower-bound of $\delta$ proposed by Ranjan et al. (2011) can be used.

Alternatively, one can consider approximating $R_n^{-1}$ by $R_{n,\delta}^{*-1} = (R + \delta J)^{-1}$, where $J$ is an $n \times n$ matrix of all 1's. A quick calculation reveals that the predicted surrogate similar to (4) - (5) will be an interpolator, unlike the scenario when we used $(R + \delta I)^{-1}$ as an approximation of $R^{-1}$. However, a thorough investigation is required to compare the model properties between $(R + \delta I)$ versus $(R + \delta J)$ approaches.

### 3.3 Reparameterisation of Correlation Functions

The estimation of the correlation hyperparameter $\theta = (\theta_1, \ldots, \theta_d)$ is the most crucial part of the GP model fitting procedure. Recall that the deviance function has to be minimized with respect to $\theta \in (0, \infty)$. For many applications, the deviance functions for such GP models are not easy to minimize. As an example, Figure 2 shows the deviance function with respect to $\theta$ for the 1-dimensional test function displayed in Figure 1. Since a large value of $\theta$ implies wigglier surrogate, we do not expect the estimated $\theta$ to be too big here. As a result, the deviance is a non-trivial function to minimize.

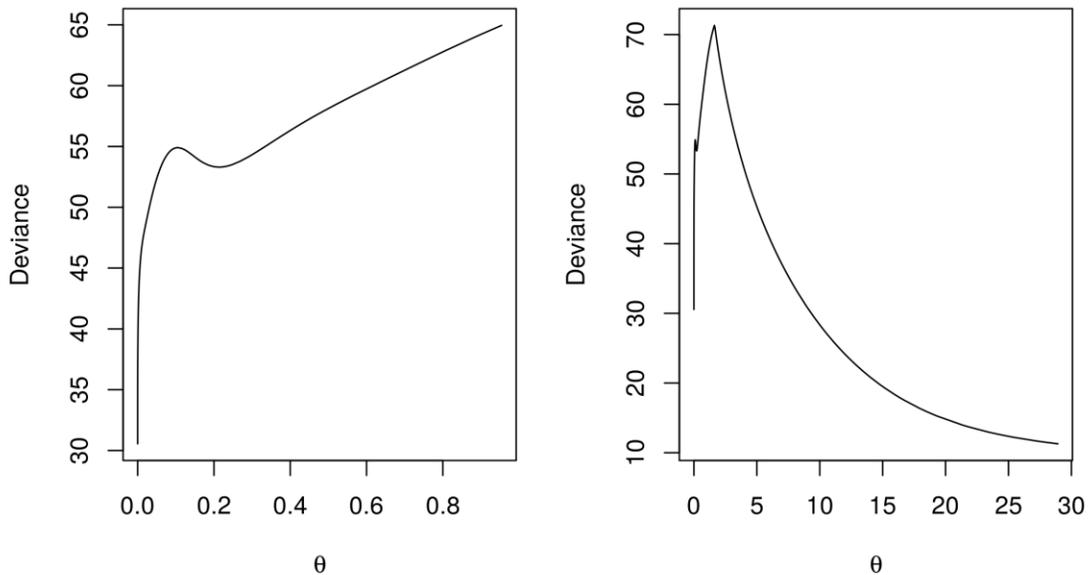

*Figure 2 Deviance with respect to $\theta$ for the test function and data used in Figure 1. The left panel is the zoomed-in version of the right panel near zero.*

In the computer experiment literature, researchers have considered a variety of re-parametrizations of this Gaussian correlation. In this section, we briefly discuss these parametrizations and compare their suitability for easier optimization.

A popular alternative representation of the correlation function uses $\lambda_k$ $(= 1/\theta_k)$, and refers to it as a correlation length parameter (Santner et al., 2013). Thus, the correlation function becomes:

$$R(x_i, x_j) = \exp\{-\sum_{k=1}^{d} |x_{ik} - x_{jk}|^2 / \lambda_k\},$$

where $\lambda_k \in (0, \infty)$. Of course, this correlation length parameter has more intuitive interpretation, and $\lambda_k$ close to zero indicates low spatial correlation and large $\lambda_k$ implies high correlation between $y(x_i)$ and $y(x_j)$. However, as expected, this parameterization would not really ease of the optmization of likelihood with respect to $\lambda_k$.

Linkletter et al. (2006) replaced $\theta_k$ with $-4\log(\rho_k)$, i.e., the new correlation hyperparameter, $\rho_k \in (0,1)$. This parametrization gives slightly better interpretability, as $\rho_k$ close to 1 means smoother fit with highly correlated nearby responses, whereas $\rho_k$ close to zero indicates spatially uncorrelated (i.e., very wiggly) surrogate fit. Unfortunately, this parametrization does not help much in the optimization of likelihood with respect to $\rho_k$. For the same 1-dimensional test function and data as used in Figure 2, the deviance surface with respect to $\rho$ is equally difficult to optimize.

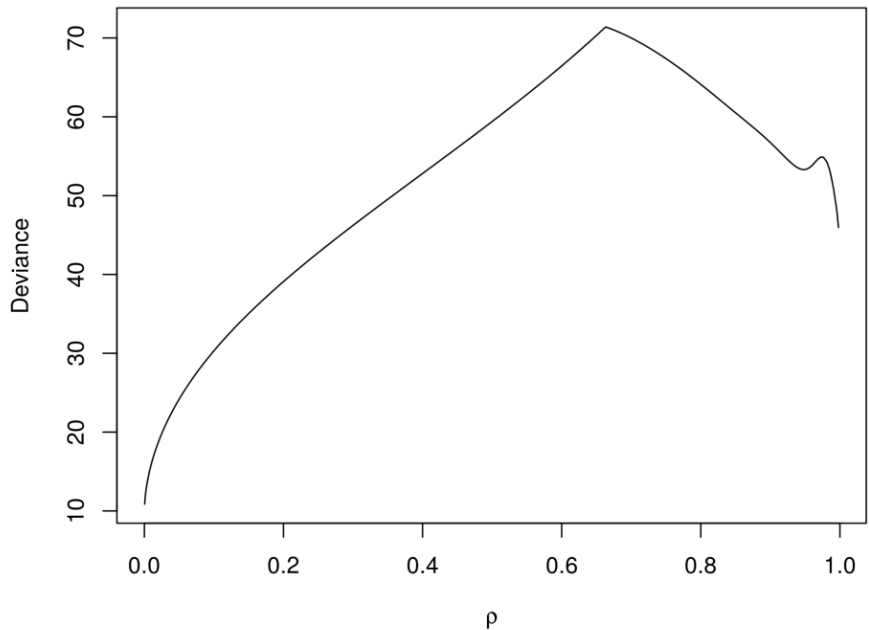

*Figure 3 Deviance with respect to ρ for the test function and data in Figure 1.*

Recently, MacDonald et al. (2015) suggested another parametrization using $\beta_k$ $(= \log_{10}(\theta_k))$. The main idea here is that the search for smooth fits correspond to negative $\beta_k$ values, whereas, wigglier surrogates are represented by large positive $\beta_k$. Moreover, the search space is now linearized, so the optimization would be lot easier. Figure 4 presents the likelihood function with respect to $\beta$, and clearly this is a better function to minimize as compared to other parametrization presented above.

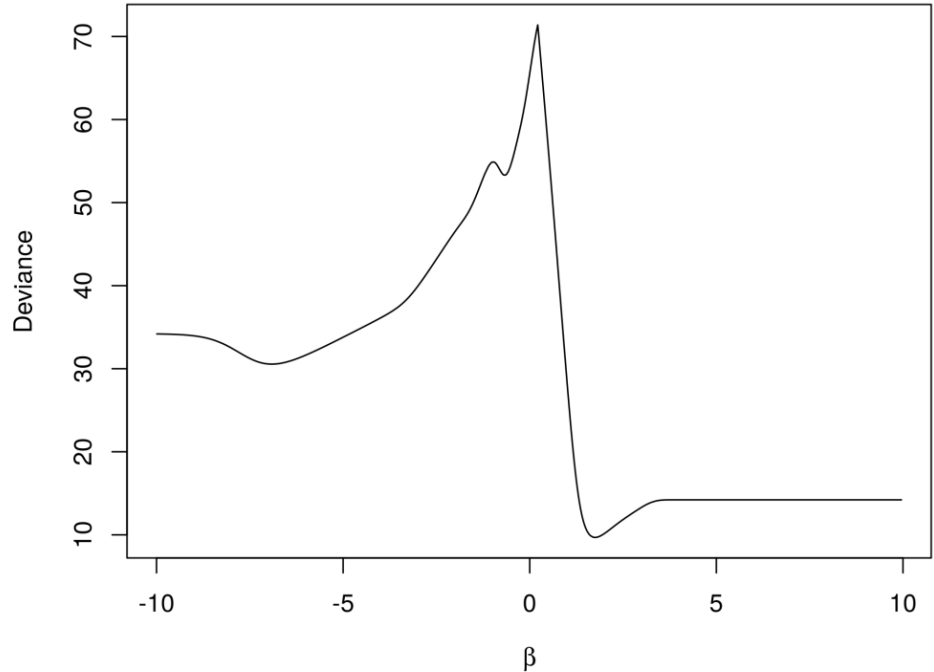

Figure 4 Deviance with respect to β for the test function and data in Figure 1.

It is important to note that the practitioners have the liberty to choose an alternative correlation structure all together instead of Gaussian correlation. However, reparametrizations discussed above can also be applied to another correlation structure.

### 3.4 Choice of Correlation Function

Historically, Gaussian correlation function or kernel is the most popular choice for defining spatial correlation in many stochastic processes. The applications range from Geostatistics to Machine Learning and Artificial Intelligence. The commonly used related terminologies are kriging and radial basis kernel.

Recall that the Gaussian correlation is a special case of the power exponential correlation function presented in (2). For real-life applications the power parameters $p_k \in [1,2]$, which can also be estimated along with other model parameters. Assuming the other model parameters are fixed, $p_k$ controls the smoothness (differentiability) of the predicted surrogate surface. See Figure 5 for an illustration of the GP model with different power exponential correlation for the same 1-dimensional test function as in in Figure 1.

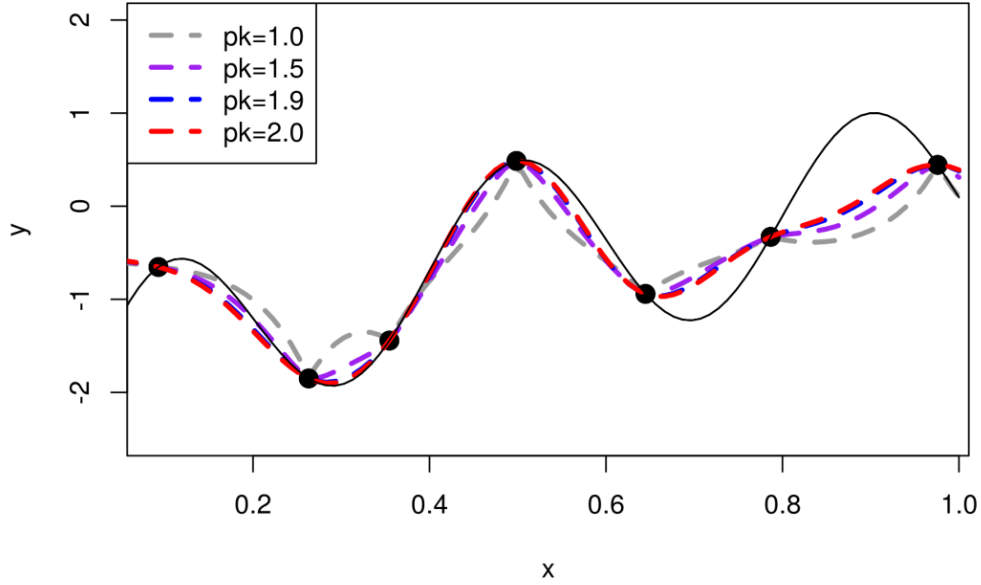

*Figure 5 Mean prediction as per the GP model fit with power exponential correlation with different $p_k$ for the test function and data used in Figure 1. The solid curve represents the true underlying simulator output.*

Figure 5 shows that the predicted surrogate is spikier at the training points as $p_k$ gets closer to 1, and much smoother as $p_k$ is closer to 2. Theoretically, it can be shown that for $p_k \in [1,2)$, the correlation kernel is differentiable only once, whereas for $p_k = 2$, the kernel is infinitely differentiable. Thus, the Gaussian correlation may seems like the most desirable correlation kernel for GP modelling, however, as shown in Ranjan et al. (2011), the probability of a correlation matrix being ill-conditioned is substantially reduced if the power is reduced from $p_k = 2$ to even $p_k = 1.95$. Furthermore, from a practical standpoint, $p_k$ close to 2 leads to reasonably smooth predictor (see $p_k = 1.9$ vs. $p_k = 2.0$ curves in Figure 5).

Another popular correlation kernel, originated from the kriging literature in Geostatistics, is called the Matern correlation. This correlation was originally obtained by letting the parameter in the Gaussian correlation follow Gamma distribution which yielded a positive and spherically symmetric density proportional to $R(x_i, x_j)$ and then finding that its Fourier transform was also a probability density (Guttorp and Gneiting, 2006). The Matern correlation kernel is given by:

$$R(x_i, x_j) = \prod_{k=1}^{d} \frac{1}{\Gamma(\nu)2^{\nu-1}} (\sqrt{2\nu}|x_{ij} - x_{jk}|\theta_k)^{\nu} \kappa_\nu(\sqrt{2\nu}|x_{ij} - x_{jk}|\theta_k), \qquad (9)$$

where $\kappa_\nu$ is modified Bessel function of order $\nu$ and $\Gamma(n)$ is Gamma function calculated at $n$. The sample paths are $\lceil \nu \rceil - 1/2$ times differentiable.

For large datasets in particular, Kaufman et al. (2011) used compactly supported correlation kernel to make the correlation matrices *sparse*, which leads to efficient evaluation and hence optimization of the likelihood using sparse matrix algorithms. Let $\tau = (\tau_1, \ldots, \tau_d)$ be the cutoff to determine the trimmed support of the design points, such that, $R_k(|x_{ik} - x_{jk}|; \tau_k) = 0$ whenever $|x_{ik} - x_{jk}| \geq \tau_k$, where $R_k(x_{ik}, x_{jk})$ represents the correlation between $x_i$ and $x_j$ for the $k$-th coordinate. Assuming the product correlation form as earlier, let

$$R(x_i, x_j; \tau) = \prod_{k=1}^{d} R_k(|x_{ik} - x_{jk}|; \tau_k),$$

where Kaufman et al. (2011) used $R_k(h_k; \tau_k) = (1 - h_k/\tau_k)\cos(\pi h_k/\tau_k) + \sin(\pi h_k/\tau_k)/\pi$. This correlation kernel is twice differentiable and is mean square differentiable.

The range parameter, $\tau_k$, plays an important role in this approach; very similar but greater than the role of $\theta$ in power exponential correlation. First, they control the degree of correlation in each dimension like correlation hyperparameter, $\theta$. Second, unlike $\theta_k$, $\tau_k$ controls the degree of sparsity in the matrix.

## 4. VARIATIONS OF GP MODELS

The GP model described thus far is the most basic version of the statistical surrogate developed by Sacks et al. (1989) for emulating the outputs of a scalar-valued deterministic computer model. Over the period of time, a host of variations and generalizations have been developed. In this section, we briefly review a few popular generalizations.

### 4.1 Non-constant Mean Function

In the context of GP models with different mean functions, thus far, four different types of Kriging models have been developed: Ordinary Kriging, Simple Kriging, Universal Kriging and Blind Kriging. The GP model presented in Section 2 is referred to as the Ordinary Kriging model (i.e., the model with constant mean $\mu$).

If we pre-specify $\mu = 0$ in the GP model of Section 2, it is referred to as the *Simple Kriging*. The closed form expressions for $\sigma_z^2$ and the mean prediction along with the uncertainty estimates are obtained by substituting $\mu = 0$ in the expressions for Ordinary Kriging:

$$E(y(x_0)|Y) = r(x_0)'R_n^{-1}Y, \quad Var(y(x_0)|Y) = \sigma_z^2(1 - r(x_0)'R_n^{-1}r(x_0)),$$

where $\hat{\sigma}_z^2 = Y'R_n^{-1}Y/n$ and the correlation hyperparameter $\theta$ (or another equivalent parameter) is estimated by maximizing the profiled likelihood.

*Universal Kriging* is a generalization of the Ordinary Kriging model, with the mean term $\mu$ being a linear function of the known basis, i.e., $\mu(x_0) = \sum_{j=0}^{m} f_j(x_0)\gamma_j$, where $\gamma_0$ is typically an intercept like term with $f_0(x_0) = 1$ for all $x_0$. The parameters and the mean prediction are obtained similarly as in the Ordinary Kriging, i.e.,

$$\hat{\gamma} = (F'R_n^{-1}F)^{-1}(F'R_n^{-1}Y), \quad \hat{\sigma}_z^2 = \frac{(Y-F\gamma)'R_n^{-1}(Y-F\gamma)}{n},$$

and

$$E(y(x_0)|Y) = f(x_0)'\gamma + r(x_0)'^{R_n^{-1}}(Y - F\gamma), \quad Var(y(x_0)|Y) = \sigma_z^2(1 - r'(x_0)R_n^{-1}r(x_0)).$$

Since it is impractical to assume that the basis functions in the mean term are known beforehand, Joseph et al. (2008) developed a new methodology to choose an appropriate set of basis functions from a class of feasible bases, for the problem at hand. They referred to this variation as the *Blind Kriging* model.

## 4.2 Noisy GP Model

As discussed earlier, realistic simulators of complex processes are sometimes non-deterministic, and hence the GP models presented thus far are not very appropriate to emulate such simulator behaviour. In the Machine Learning and Computer Experiment literature, the following version of the GP model has gained much popularity:

$$y_i = \mu + z(x_i) + \varepsilon_i, \quad i = 1,2,..,n,$$

where the additional error term $\varepsilon_i$'s are iid $N(0, \sigma_\varepsilon^2)$ and independent of $\{z(x), x \in [0,1]^d\}$, the GP with mean zero, variance $\sigma_z^2$ and correlation kernel $R(\cdot,\cdot)$, as defined earlier (see Santner et al. (2003) for details). Of course, one can use different mean function instead of a constant mean $\mu$ as discussed in the previous section.

The inclusion of an additional error term does not introduce much deviation from the regular model fitting procedure, because the joint distribution of $Y = (y_1, y_2, \ldots, y_n)$ is multivariate normal with mean $\mu 1_n$ and variance-covariance matrix $\Sigma = \sigma_z^2 R_n + \sigma_\varepsilon^2 I_n$, where $I_n$ is the $n \times n$ identity matrix. Note that rewriting $\Sigma = \sigma_z^2(R_n + \delta I_n)$, where $\delta = \sigma_\varepsilon^2/\sigma_z^2$ translates this model to the GP model with a nugget term as in Ranjan et al. (2011). Of course, here $\delta$ will also have to be estimated along with other model parameters. As earlier, one must be cautious in defining the search space for $\delta$ as very small $\delta$ may lead to near-singular/ill-conditioned $\Sigma$. Moreover, there is no need to adopt the iterative regularization as the simulator is noisy and interpolation is not the objective.

## 4.3 Dynamic GP Model

Higdon et al. (2008) proposed an SVD-based GP model for the emulation of computer simulators with highly multivariate outputs, and recently, Zhang et al. (2018b) used it for simulators with time series responses. Consider a deterministic simulator with $d$-dimensional input $\mathbf{x} \in \mathbb{R}^q$, which returns a time series output $\mathbf{y}(\mathbf{x}) \in \mathbb{R}^L$ of length $L$.

Let $\mathbf{X} = [\mathbf{x}_1, \ldots, \mathbf{x}_N]^T$ be the $N \times q$ input matrix and $\mathbf{Y} = [\mathbf{y}(\mathbf{x}_1), \ldots, \mathbf{y}(\mathbf{x}_N)]$ be the $L \times N$ matrix of time series responses, then the SVD on $\mathbf{Y}$ gives $\mathbf{Y} = \mathbf{U}\mathbf{D}\mathbf{V}^T$, where $\mathbf{U} = [\mathbf{u}_1, \ldots, \mathbf{u}_k]$ is an $L \times k$ column-orthogonal matrix of left singular vectors, with $k = min\{N, L\}$, $\mathbf{D} = \text{diag}(d_1, \ldots, d_k)$ is a $k \times k$ diagonal matrix of singular values sorted in decreasing order, and the matrix $\mathbf{V}$ is an $N \times k$ column-orthogonal matrix of right singular vectors. The SVD-based GP model for a deterministic simulator is given by,

$$\mathbf{y}(\mathbf{x}) = \sum_{i=1}^{p} c_i(\mathbf{x})\mathbf{b}_i + \varepsilon, \tag{10}$$

where the orthogonal basis $\mathbf{b}_i = d_i \mathbf{u}_i \in \mathbb{R}^L$, for $i = 1, \ldots, p$, are the first $p$ vectors of $\mathbf{U}$ scaled by the corresponding singular values. The coefficients $c_i$'s in (10) are random functions assumed to be independent scalar response GP models, i.e., $c_i \sim \text{GP}(0, \sigma_i^2 R_i(\cdot, \cdot; i))$ for $i = 1, \ldots, p$ (Rasmussen and Williams, 2006). The residual error $\varepsilon$ in (10) is assumed to be independent Gaussian white noise, that is, $\varepsilon \sim \mathcal{N}(0, \sigma^2{}_L)$.

The built-in function called svdGP in the R package DynamicGP provides an easy implementation of this surrogate model (Zhang et al., 2018a). The arguments of svdGP can also be tuned to speed up the computation by parallelization.

## 4.4 Non-stationary GP Model

Though we have not been very explicit yet, most of the discussion on GP models assumed that the underlying process / phenomena is stationary. The standard GP itself is defined to be covariance (i.e., weak) stationary. However, in reality, there are several phenomena that are non-stationary, which in a lay man terms is like a function with abrupt changes in the curvature or shape. For instance, Figure 6 shows two real-life applications. The left panel represents the output of a simplified simulator which generates the average maximum extractable power from the Minas Passage, Bay of Fundy, Nova Scotia, Canada, given that one turbine fence is already present in the Passage (Chipman et al., 2012). The right panel presents the simulated measurements of the acceleration of the head of a motorcycle rider as a function of time in the first moments after an impact (see mcycle data in the R library MASS for details). These are undoubtedly non-stationary processes, and standard GP models would not serve as adequate surrogate models (see the rightmost panel of Figure 7).

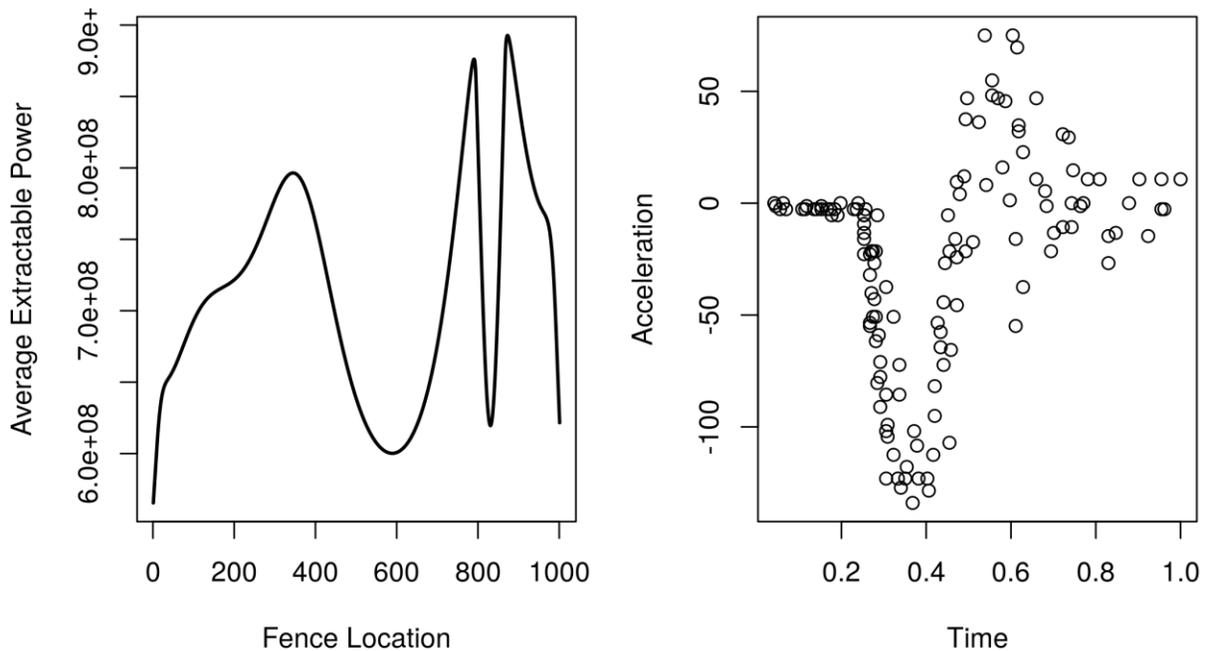

*Figure 6 Left panel: maximum extractable power from the Minas Passage (simplified computer model for turbine placement in the Bay of Fundy, Nova Scotia, Canada, see Chipman et al. (2012)). Right panel: acceleration of the head of a motorcycle rider as a function of time in the first moments after an impact (see `mcycle` in R library MASS for details).*

One naive way to capture the non-stationarity is to detrend the data via carefully chosen mean basis (as discussed in Section 4.1), and then use the standard GP model to emulate the residual stationary process. Over the last two decades, several innovative surrogates have also been developed to emulate the non-stationary computer model responses. For instance, Higdon et al. (1999) made some fundamental methodological contribution towards the non-stationary correlation structure, but the computer experiment literature itself was not mature enough until early - mid 2000's. Paciorek and Schervish (2004) further formalized this GP-based emulator. Gramacy (2007) combined the idea of regression trees with GP model and developed Treed GP model (TGP), which is simply fitting GP models instead of constants to the terminal nodes. Ba and Joseph (2012) suggested a sum of two GP model strategy to separately capture the local nuances and fluctuations versus the overall global trend. Chipman et al. (2012) further demonstrated that a Bayesian Additive Regression Tree (BART) can easily be used to emulate non-stationary computer simulator outputs and are perhaps more reliable than many other competitors for large datasets (see Figure 7 for an illustration on the motorcycle data). Recently, Volodina and Williamson (2018) used a mixture of GP based approach for this surrogate building exercise.

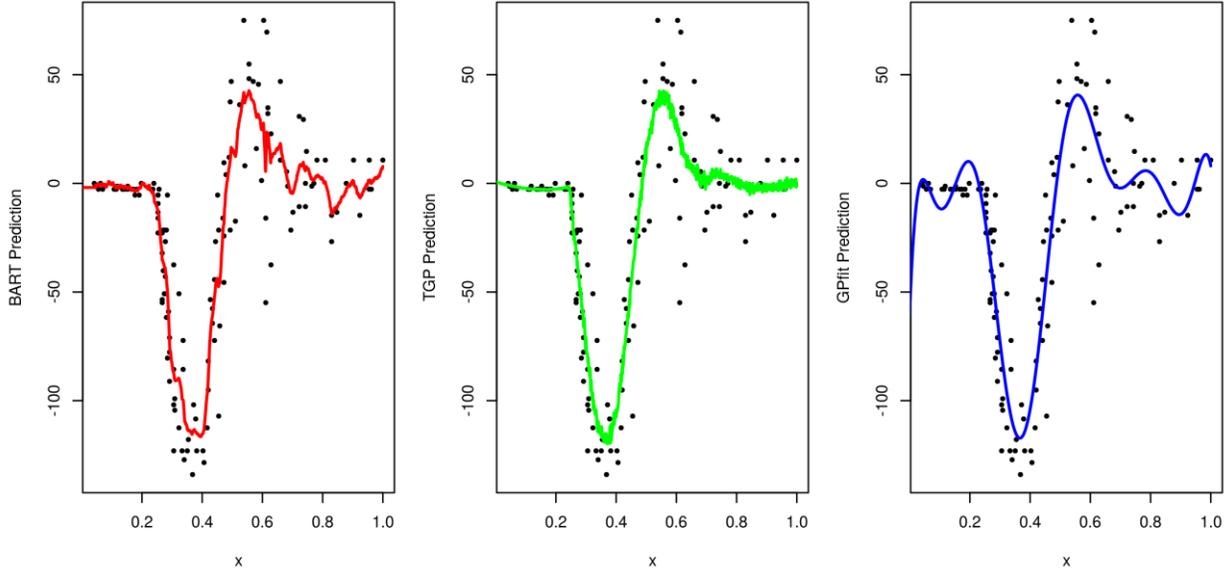

*Figure 7 Mean prediction for the Motor Cycle data (source: R library MASS) as per BART model (Chipman et al., 2012), TGP model (Gramacy, 2007) and standard GP model via GPfit (MacDonald et al., 2015) (in order, from left to right).*

## 5. BIG DATA AND HIGH PERFORMANCE COMPUTING

The accelerated growth in the computing power of data processing and storage has led to a new area of science called the BIG data. Specifically, the data from computer simulators can easily get really large if the simulator is computationally fast. Over the last decade, the researchers have been investigating both aspects, the innovative methodologies for modelling and analysis, and efficient implementation techniques and algorithms for BIG data obtained from computer simulators.

### 5.1 Methodological Innovations

For Gaussian process models, exact calculations of $R_N^{-1}$ requires $O(N^3)$ operations, which has to be done numerous times for likelihood optimization. Thus, efficient evaluation of the likelihood function is extremely crucial for GP modelling for a very large training dataset (of size $N$, say). Though there are several interesting methodological contributions, we briefly discuss a few very popular ones.

Stein et al. (2004) proposed an approach to break down the joint multivariate normal density into a product of conditional densities that significantly reduces the computational time. Furrer et al. (2006) and Kaufman et al. (2011) suggested using "tapering" in the correlation matrices via a compactly supported kernel (see Section 3.4), so that the sparse matrix algorithms can be better utilised for computational savings.

Another line of approach is to replace one big common GP model on a very large dataset

(say $N$) with several local models based on small datasets of (say) size $n$ ($\ll N$) each for approximating the predicted response at an arbitrary $\mathbf{x}_0$ in the input space. Let $\mathbf{X}$ be the large training set of $N$ points, and $\mathbf{X}^{(n)}(\mathbf{x}_0)$ or $\mathbf{X}^{(n)}$ (in short) denote the desired subset of which defines the $n$-point neighborhood of $\mathbf{x}_0$ contained in $\mathbf{X}$. We briefly discuss two methods of constructing this neighborhood set $\mathbf{X}^{(n)}$. The first one, called as the *naive* approach, assumes the elements of the neighborhood set $\mathbf{X}^{(n)}$ by finding $n$ nearest neighbors of $\mathbf{x}_0$ in $\mathbf{X}$ as per the Euclidean distance in the *k-nearest neighbor* method. The emulator obtained via fitting a GP model to this local set of points is referred to as *k-nearest neighbor GP model* (in short, knnGP). Though, knnGP is computationally much cheaper than the *full GP model* (in short, fullGP) trained on $N$ points, its prediction accuracy may not be satisfactory. Emery (2009) finds the neighborhood set $\mathbf{X}^{(n)}(\mathbf{x}_0)$ (for every $\mathbf{x}_0$) using a greedy approach. Gramacy and Apley (2015) further improved the prediction accuracy by using a sequential greedy algorithm and an optimality criterion for finding a non-trivial local neighborhood set (see Figure 8 for an illustration). This method is also tailored for computation on modern day multi-processing, multi-threaded computers.

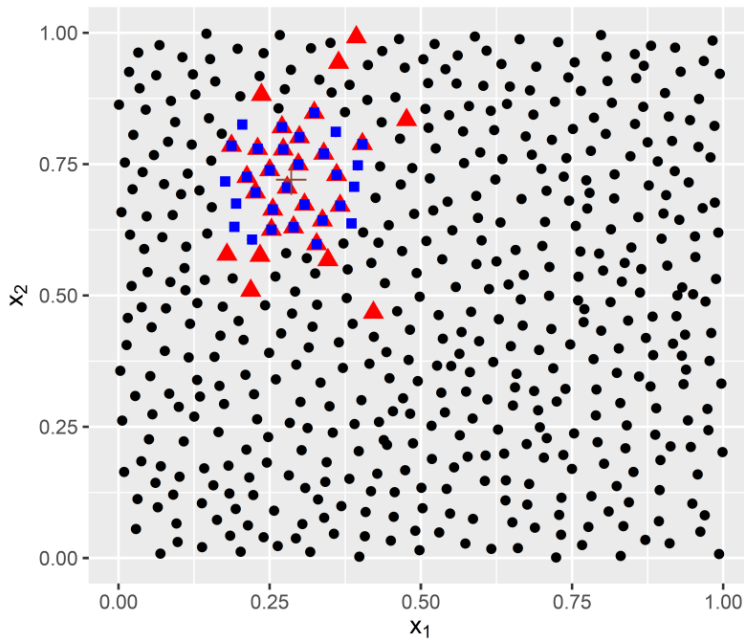

*Figure 8 Local neighbourhood selection as per the k-nearest neighbour method (blue squares) and the greedy approach (red triangle) by Gramacy and Apley (2015) for prediction at the location marked by black plus.*

## 5.2 Computational Efficiency

In recent times, researchers have started focussing on the development of algorithms that are computationally efficient, can easily be parallelized, and in particular suitable for large data sets. Many of the software packages that are now being released, come with MPI, Open MP and CUDA code components, which have the option of running codes in parallel and/or use the built-in GPU components for faster processing.

Despite using sophisticated methodologies developed for handling big data, fitting GP models for big data can often be computationally expensive. Franey et al. (2012) demonstrate how Graphics Processing Units (GPU) give us more computing power than Central Processing Units (CPU) for standard GP models. For a quick reference, Table 1 presents a comparison of computation time for the standard CPU computing versus CPU+GPU implementation. Note that the results were obtained on a naive high performance computing (HPC) supported desktop (that a student could afford in 2011, the time of research), and now a much more significant improvement can be recoded on the latest HPC platform.

*Table 1 Performance comparison of standard GP model fits. The outputs are generated via Hartman-6 function, and the inputs are random maximin Latin hypercube designs in [0,1]^6. The results are averaged over 10 simulations, except the last row of CPU implementation (denoted by ∗), which is based on only 1 simulation. See Franey et al. (2012) for details.*

| CPU Implementation | | | | | |
|---|---|---|---|---|---|
| $n$ | Time $(s)$ | $-2logL_\theta$ | $\hat{\mu}$ | $\hat{\sigma}_z^2$ | SSPE |
| 64 | 32.32 | 125.94 | 0.1771 | 0.1403 | 77.4160 |
| 256 | 514.43 | 610.25 | 0.1105 | 0.1164 | 27.4311 |
| 1024 | 13325.86 | 2491.97 | 0.0609 | 0.0970 | 5.6504 |
| 4064 | *161925.05 | *8044.80 | *0.0485 | *0.0824 | *0.5320 |
| CPU Implementation | | | | | |
| $n$ | Time $(s)$ | $-2logL_\theta$ | $\hat{\mu}$ | $\hat{\sigma}_z^2$ | SSPE |
| 64 | 9.45 | 103.70 | 0.1238 | 0.2989 | 91.7860 |
| 256 | 16.58 | 547.10 | 0.1397 | 0.1746 | 31.4641 |
| 1024 | 96.19 | 2665.58 | 0.1192 | 0.1390 | 4.3850 |
| 4064 | 1059.71 | 8698.28 | 0.0803 | 0.0700 | 0.5314 |

In summary, parallel-running GPUs when combined with CPUs are far more effective on per-dollar basis than most multi-core CPUs (alone). Gramacy et al. (2014) and Liu et al. (2018) investigated it further and developed more advanced methodologies and implementation algorithms particularly advantageous for large data sets. New R libraries like laGP (Gramacy, 2015) and DynamicGP (Zhang et al., 2018a) takes the advantage of multi-core processors and run specific tasks in parallel. One can also specify the number of threads to be assigned for a particular code in these packages.

HPC on Microsoft R has recently been gaining popularity as well. Microsoft R is an enhanced version of R which supports multithreading for calculations. The original R was designed to use single thread for computations and this modified version adds *Intel Math Kernel Library* (IMKL) which significantly decreases computational expenses. Microsoft R functions exactly like R; so there is no change required in the code or library. Matrix operations in particular are immensely benefited by using multithreading approach. The benchmark reports can be accessed at https://mran.microsoft.com/documents/rro/multithread. To reproduce the results and better understanding, one can see GitHub repository: https://github.com/andrie/version.compare.

## 6. DATA ANALYSIS GOALS

There are several popular pre-specified objectives of running computer simulators and data analysis. For instance, (a) the overall understanding of the entire simulator response surface, (b) the estimation of a predetermined feature of interest, such as, the global minimum, a contour (also referred to as the inverse solution), a quantile, etc. (c) the calibration of the simulator itself, and (d) identification of important input variables.

A major portion of computer experiment literature emphasize on the "*design of computer experiments*", which refers to the technique of choosing a set of input combinations ($x$'s) for running the computer simulator. For objective (a) listed above, several good designs have been developed. One of the most popular jargon in this section of the literature is Latin hypercube based designs with space-filling properties like maximin interpoint distance, minimum pairwise-coordinate correlation, and so on. Given that the goal is to explore the overall simulator response surface, the most common form of analysis is the "sensitivity analysis" - which sort of overlaps with objective (d).

Over the last two decades, many researchers in this area have focussed on developing innovative methods and algorithms for estimating process optimum. However, this was under the assumption that the computer simulator is computationally expensive to run, and subsequently, the training data is not large enough to be classified as BIG data. Though the size of the training data can sometimes be in thousands, the corresponding input dimension is too large (e.g., 20) to prevent thorough exploration of the input space using 1000 points. That is, the budget for the total number of simulator runs (say $n$) is pre-fixed and too small (with respect to the input dimension $d$) to use traditional optimization techniques, which led to the need for a new method for global optimization in computer experiments.

Jones et al. (1998) proposed an efficient sequential design scheme for finding the global minimum. The algorithm starts with choosing an initial design of size $n_0 (\ll n)$ and then selects the remaining $n - n_0$ follow-up points sequentially one at-a-time by maximizing a merit-based criterion called the *expected improvement* (EI). In Jones et al. (1998), the EI criterion is simply the expectation of the improvement function,

$$I(x) = \max\{f_{min}^{(k)} - y(x), 0\},$$

with respect to the predictive distribution of $y(x)$ given the observed data on $n_0 + k$ points, where $f_{min}^{(k)}$ is the running estimate of the global minimum, and $y(x)$ is the unobserved response at the input $x$. This approach gained significant popularity because the EI criterion facilitated a tradeoff between the local exploitation and the global exploration, i.e., the algorithm made sure that the global minimum was found and did not get stuck in the local optimum. Since then a plethora of slightly different EI criteria have been proposed for the global minimum (see, for example, Schonlau et al. (1998) and Santner et al. (2003)).

Ranjan et al. (2008) extended the EI approach for estimating a pre-specified contour (also popularly referred to as the inverse solution) from an expensive to evaluate deterministic computer simulator. The notion of contour estimation was further adopted for quantile estimation and estimating the tail probability of failure (see Bingham et al. (2014) for a review). The complexity of the estimation of an inverse solution increased substantially when the computer simulator returns a time-series response instead of a scalar. Ranjan et al. (2016) tried to use a standard GP model based EI approach via scalarization technique for solving this inverse problem. Vernon et al. (2010b) proposed a history matching algorithm for this purpose, and recently, Zhang et al. (2018c) further extended the EI approach under the SVD-based GP models for dynamic simulator response.

## 7. REAL-LIFE COMPUTER MODELS

The applications of real-life computer models range over a wide spectrum of discipline, from behavioural models to the simulation of a nuclear reaction. In this section, we present brief descriptions of a few real-life simulators.

**TDB model:** The underlying objective is to gain a thorough understanding of the population growth of a pest called the European red mites (ERM) or Panonychus ulmi (Koch). ERM infest on apple leaves, resulting in poor yields, and hence a concern for apple farmers in the Annapolis Valley, NS, Canada. Franklin (2014) developed a mathematical-biological model based on predator-prey dynamics called the Two-Delay Blowfly (TDB) model, which consists of eleven parameters treated as the inputs to the model, and produces time-series outputs that characterize the ERM population growth. Ranjan et al. (2016) did some preliminary research on the calibration of this simulator. Recently, Zhang et al. (2018b) built a dynamic GP model for analyzing BIG data obtained from the TDB model. Zhang et al. (2018c) further extended the work to find the optimal set of inputs of the TDB model that gives a good approximation to a pre-specified target (e.g., the field data). The intent behind this inverse problem was to calibrate the TDB model to produce realistic outputs closer to the reality.

**Tidal power model:** The Bay of Fundy, located between New Brunswick and Nova Scotia, Canada, is world famous for its high tides. Among others, Karsten et al. (2008) suggested harnessing this green / renewable tidal energy by installing a host of in-stream tidal turbines. However, the cost of building a turbine and installing it in the Bay of Fundy is extremely high (in millions of dollars). Thus, it is desirable to minimise the number of turbines to harness the maximum extractable total power. However, a physical experiment to find the optimal locations of these tidal turbines is infeasible due to the cost constraint. Karsten et al. (2008) developed a version of the finite volume community ocean model (Greenberg, 1979), for preliminary analysis and experimentation in the Minas Passage of the Bay of Fundy. Ranjan et al. (2011) used this computer model for finding the optimal location of one tidal turbine by maximising the power function. Chipman et al. (2012) used non-stationary surrogate model based optimization strategy for finding the optimal locations of several turbine fences for a case study by Karsten et al. (2008).

**SWAT model:** Soil and Water Assessment Tool (SWAT) model is an internationally recognised computer model which simulates runoff from watershed areas based on climate variables, soil types, elevation and land use data (Arnold et al, 1994). Bhattacharjee et al. (2017) used a modified history matching algorithm built upon the GP-based surrogate for calibrating this model with respect to the Middle Oconee River (Georgia, USA) data. The idea of history matching was popularised by Vernon et al. (2010, 2014) when calibrating a Galaxy formation simulator called GALFORM.

**MRST model:** Finding the optimal drilling locations for production and injection wells in an oil reservoir is of utmost importance (see Onwunalu and Durlofsky, 2010). Butler et al. (2014) used a Matlab Reservoir Simulator (MRST) (Lie et al., 2012) to generate the anticipated net present value (NPV) of the produced oil for a well to be drilled at a particular location. The goal here was to determine the configuration of wells that yields the best NPV.

## 8. CONCLUSION AND FUTURE DIRECTIONS

In this chapter, we have reviewed the Machine Learning and Statistics literature on design, analysis and modelling of data arising from computer simulation models. Recall that computer models are often used as cheaper alternatives for complex physical phenomena, however, simulators can also be computationally too expensive for thorough experimentation, and for which, statistical models are used to emulate the simulator output. For the last two decades, realisations of Gaussian process (GP) models are used for this emulation. Section 2 of this chapter presented a brief review of the most basic GP regression model. This non-linear semi-parametric regression model may appear to be straightforward, however, the numerical issues in fitting this model are somewhat involved. In Section 3, we have briefly reviewed the major computational concerns, i.e., the near-singularity of the correlation matrix, efficient matrix decomposition methods, choice of correlation kernels, and the reparametrization of the correlation length parameters. Section 4 summarized a variety of popular GP-based surrogates under the generalised setup such as non-stationarity, dynamic response model, and stochastic simulators. The treatment of BIG data obtained from computer models was briefly discussed in Section 5, and Section 6 reviewed a few popular analysis goals of such computer experiments. Finally, Section 7 presented a brief description of a few real-life computer models.

With respect to the future research directions, the relentless growth in the computing power demands for more advanced methodologies and efficient algorithms. Furthermore, not all methodologies developed thus far are full proof in every aspect. For instance, the nugget based approach had been developed only when the GP model was to be built for the overall good fit. If the objective is to estimate a pre-specified feature of interest like the global optimum, then the proposed lower bound of the nugget would not work, and is still an open research problem. On the other hand, the power-exponential correlation with $\boldsymbol{p_k < 2}$ (say 1.95) substantially reduces the chances of near-singularity, however, does not completely resolves it. The development of new methodologies and analysis for dynamic GP models is still at the early stage, and much further work have to be done, e.g., the construction of optimal design for different analysis objectives. Under the umbrella of BIG data, most of the innovative work thus far focus on tweaking the existing methodologies, for example, via conditional likelihood or sparse computations. New innovative methodologies and algorithms (e.g., for building specific surrogate model and constructing optimal design) tailored towards BIG data are still open research problems.

## ACKNOWLEDGMENT

The authors would like to thank the Editor and four anonymous referees for their thorough and helpful reviews. Ranjan's research was partially supported by the Extra Mural Research Fund (EMR/2016/003332/MS) from the Science and Engineering Research Board, Dept. of Science and Technology, Govt. of India.